\documentclass[%
 aip,jmp,amsmath,amssymb,reprint
]{revtex4-2}

\usepackage{datetime}
\usepackage[version=4]{mhchem}
\usepackage{times,mathptmx}
\usepackage{graphicx}
\usepackage{wrapfig}
\usepackage{dcolumn}
\usepackage{eucal}
\usepackage{relsize}
\usepackage{physics}
\usepackage{sidecap}

\def\r{\right}
\def\l{\left}

\def\half{{\textstyle\frac{1}{2}}}
\def\2D{\mathsf{2D}}

\begin{document}

\title[working draft]{
Quantifying `local softness' in a simple liquid
}

\author{Tamoghna Das}
\email{tamoghna@staff.kanazawa-u.ac.jp}
\affiliation{%
Nano Life Science Institute, Kanazawa University \\
Kakuma-machi, Kanazawa, Japan, 920-1192
}%
 


\begin{abstract}
Mutual information between local stress and local non-affine deformation is proposed as a collective field variable quantifying the {\em local softness} of soft materials. The liquid-solid transition in a simple liquid is considered as a generic example of mechanical transformation through varying correlation between stress and deformation at the microscopic level. Probing through this new measure, a liquid appears as a spatially heterogeneous medium of interacting interconnected regions of varying softness. In contrast, the soft regions shrink to isolated spots in the background of a negligible mean softness in the case of solids. In this view, the thermodynamic transition becomes purely geometric while keeping the essential mechanical information intact. Besides offering a general framework for understanding the mechanics of materials, this new approach can complement recent machine learning efforts by assigning physical meaning to their findings. Further, this collective variable can be used on the fly during material characterization as both of its ingredient variables are experimentally accessible.

\end{abstract}

\maketitle

\section{Introduction}

The non-equilibrium response of materials, such as growing rigidity under increasing mechanical perturbation,~\cite{Reynolds:1885, Rowe:1962, Sastry:2021} strain-rate-dependent yielding,~\cite{Thomson:1974, Wolynes:2012, Tarjus:2018, Sengupta:2020} shooting viscosity,~\cite{Angell:1995, Allan:2009} generally intrigues us as they defy the predictions by idealised equilibrium theories. Such non-linear, dynamically evolving and, most importantly, system-specific responses are notoriously challenging to conceptualize within a unifying framework. They might even depend on the period of observation~\cite{Stillinger:2000, Nagel:2019} as some of the systems show aging while others might rejuvenate after a certain time. The notion of spatio-temporally evolving localized excitations have been invoked to describe such anomalies. While it is important to identify proper microscopic observables representing these excitations, it is also necessary to acquire the knowledge of full distributions of such observables. Limiting the scope to a few low-order moments, such as mean and variance of the observable, as often done in the equilibrium theories, is inadequate to capture the full complexity of the problem.


In all aforementioned out-of-equilibrium phenomena, one recognizes an intricate dependency between the local change in mechanical energy or stress and simultaneous reorganization of particle configuration leading to deformation. Over the past decades, several proposals have been put forward for an appropriate observable in terms of either energy/stress or deformation. Here, we mention the soft spots theory and the shear transformation zones (STZs) theory as representatives of these two complementary approaches. STZ theory~\cite{Langer:1998, Langer:2008, Langer:2011, Egami:2012, Langer:2015} focuses on identifying local changes in the configurational space due to instantaneous accumulation or dissipation of shear stress and attempts to surmise the global mechanical properties from the statistical behaviour of such changes with external control processes such as cooling or strain cycle. On contrary, soft spots theory~\cite{Varma:1972, Laird:1991, Reichman:2008, Ma:2014, Lerner:2019} is developed to investigate the roots of bulk mechanical response within the momentum space fluctuations of the system. Relying on the fact that anomalous dynamical modes would originate from local and instantaneous energy fluctuations, this description charts out possible conformational changes to support such modes which are called soft spots. An alternative description relying on the statistical properties of local stress tensor in relation with conformational changes including the deformation of particle itself is also developed for granular materials.~\cite{Edwards:2012, Daniels:2013, Chakraborty:2015, Chakraborty:2019} 
 
In the current study, we take an information theoretic approach as a concise representation of the full distribution of microscopic observable(s). Instead of treating the local mechanical energy and local deformation as independent variables, as in the previous instances, we quantify the degree of dependence between these two field variables in terms of their mutual information which by definition is a non-zero positive quantity. While a zero mutual information would imply complete independence between the variables, non-zero values would indicate degree of coupling between stress and deformation locally. This new collective variable reveals that such coupling is spatially heterogeneous even in an equilibrium system and allows us characterise the spatial extent of this coupling. Since we do not expect any dynamic evolution of the bulk response in equilibrium, we study the spatial statistics of local mechanical coupling across the liquid-solid transition, one canonical example of equilibrium mechanical transformation, employing this new approach. In what follows, we first describe our model system and discuss the computational details of our primary observable quantities. Next, we report their statistics across phase transition, define the collective variable and present its spatial features in detail. Finally, we conclude by discussing the implications of these results along with their possible immediate applications.

\section{Model system and measured variables}
As a test system, we choose a prototype of simple liquids: a set of mono-disperse particles of unit mass interacting pair-wise via Lennard-Jones interaction within a periodic two-dimensional ($\2D$) box. The interaction, $\phi(r)=4\epsilon\l[(\sigma/r)^{12}-(\sigma/r)^6\r]$, is defined by the energy and length scales, $\epsilon$ and $\sigma$, respectively. This also sets the timescale of the problem, $\tau=\sqrt{\sigma^2/\epsilon}$. The equation of motion of the particles are solved by the velocity Verlet algorithm with a time step $10^{-3}\tau$. The temperature $T$ of the system is measured in $\epsilon$ units assuming the Boltzmann constant to be unity and is kept constant using a {\em DPD} (dissipative particle dynamics) thermostat~\cite{Warren:1997} as implemented in LAMMPS. The thermostat is chosen so as to mimic a broader class of materials, including colloids and polymeric systems. However, the thermostat dynamics has least effect on the equilibrium properties which is our current focus. The equilibrium is ensured by the fluctuations of the total energy and thermal energy, which are of the order of $10^{-5}$ about their mean values for our model system consisting of about $10^5$ particles. The mean values were observed to remain constant for about $500\tau$ time period. After a typical equilibration run, $10^3$ configurations were saved in an interval of $\tau$ for further analysis.

\begin{SCfigure}
\includegraphics[width=0.45\textwidth, height=0.45\textheight, keepaspectratio]{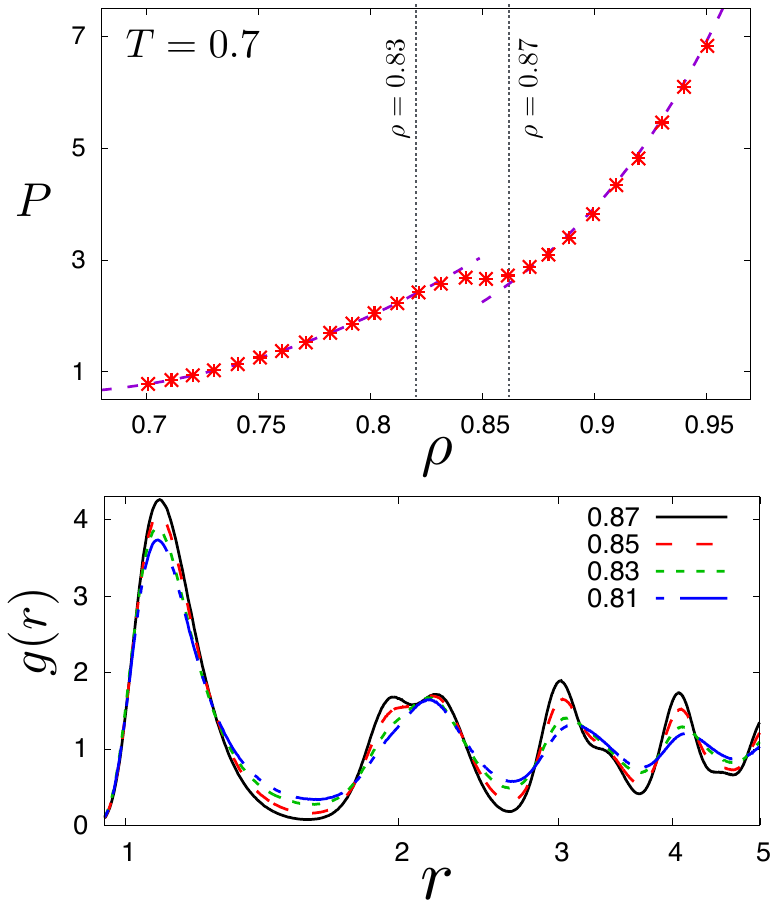}
\caption{\label{fig:basic}
{\bf Basic characterization of the model system:} (A) $P$-$\rho$ diagram of a two-dimensional single component Lennard-Jones system at $T=0.7$ showing the liquid-solid coexistence region bound between $\rho=0.83$ and $\rho=0.87$. (B) The radial distribution $g(r)$ of particles in the coexistence phase showing the appearance of peaks over the usual oscillatory feature, characteristic to the liquid at $\rho=0.81$. Particularly, note the evolution of second peak as a function of increasing density. At $\rho=0.83$, this peak becomes skewed; then at $\rho=0.85$, a shoulder appears; and at $\rho=0.87$, the peak splits into two marking the onset of crystalline phase.
}
\end{SCfigure}
At present, we are mainly concerned about two local quantities: local stress and local non-affine deformation. Below, we first define these observables along with their detailed computations. Our first observable, the local stress~\cite{Rossi:2006} is computed by the following average virial stress tensor over a volume $\Omega$ ({\em area} in our case) around a particle $i$ at position ${\bf r}_i$: 
\begin{equation}
\bar{\Pi}_i=\frac{1}{\Omega}\l[ -m_i({\bf v}_i\otimes{\bf v}_i)+\half\sum_{j\ne i}{\bf r}_{ij}\otimes{\bf f}_{ij}\r]
\end{equation}
where, ${\bf r}_{ij}$ is the distance from its $j$-th neighbour, and ${\bf f}_{ij}$ is the force exerted by the same neighbour on the $i$-th particle. Note that, ${\bf f}_{ij}=-{\bf f}_{ji}$ for the obvious reasons and $\otimes$ denotes a tensor product between two vectors. ${\bf v}_i$ and $m_i$ are the velocity and mass, respectively, of the $i$-th particle. The volume, $\Omega$, is considered to extend over a distance of $2.5\sigma$ from the $i$-th particle. As the potential practically becomes zero beyond this distance, the forces exerted by any neighbour further than this distance would be negligibly small and are neglected for the stress computation. $\bar{\Pi}_i$ would be a $(2\times2)$ matrix for our model system. In order to understand the local stress fluctuations in equilibrium, we have considered the following three scalar stress invariants.
\begin{itemize}
    \item Volumetric stress $\sigma_{v}=\half(\bar{\Pi}_i^{11}+\bar{\Pi}_i^{22})$, is experienced by a particle when there is any overall expansion or contraction of its surrounding volume $\Omega$.
    \item Deviatoric stress or {\em pure} shear, $\sigma_{d}=\half(\bar{\Pi}_i^{11}-\bar{\Pi}_i^{22})$, represents the stress due to any change in the shape of $\Omega$ which preserves its volume.
    \item {\em Simple} shear stress, $\sigma_{s}=\half(\bar{\Pi}_i^{12}+\bar{\Pi}_i^{21})$, accounts for all other types of unrestricted shape changes of $\Omega$. 
\end{itemize}
In equilibrium, only $\sigma_{v}$ contributes to the bulk stress response of a material as the ensemble average of this quantity equals to the bulk pressure of the system. The two shear stress components do not appear as any such bulk response. Nevertheless, their fluctuation statistics reveal certain interesting properties of local stress response in a system undergoing phase transition as we shall see below.

Our second observable, the non-affine deformation is originated from computational geometry, where the conformational change of a simplex due to all possible translation, rotation and permutation of its nodes is quantified by the {\em procrustean distance} between the transformed and the original polygon.~\cite{Goodall:1991, DrydenMadria} For the present scenario, permutation of the particles are, of course, not allowed. Considering the immediate neighbourhood of a particle as a closed polygon, we estimate its instantaneous deformation employing the same measure, sans its permutation component:
\begin{equation}
    \chi = \sum_{\langle j\rangle} \left[{\bf r}_{ij}(t)-{\mathbb A}{\bf r}_{ij}(0) \right]^2
\end{equation}
where ${\bf r}_{ij}(t)$ is the distance between the $i$-th particle and its $j$-th neighbour at time $t$. The matrix, $\mathbb A$, represents the minimal {\em affine} transformation required to match the distance between an arbitrarily chosen pair of particles, ${\bf r}_{ij}$ at $t=0$ (chosen arbitrarily), to the instantaneous distance ${\bf r}_{ij}(t)$ between the same pair. All such mismatches between a given particle and its immediate neighbours (denoted by $\langle\dots\rangle$) at two instants are then squared and summed to compute the non-affine deformation field $\chi$~\cite{Langer:1998} which is experimentally accessible~\cite{Behringer:2008, Schall:2011} via any particle resolved imaging technique. Note that identification of the neighborhood is crucial for this computation. We have employed a geometric and non-parametric algorithm, {\em relative angular distance} (RAD),~\cite{Henchman:2016} for this purpose. Starting from the nearest neighbour, RAD algorithm collects all the distant neighbours as long as no one neighbour is blocked by any of the already existing neighbours. This method is not space-filling as the more commonly used Voronoi tessellation. However, this method is much more computationally efficient and well suited for inhomogeneous arrangement of particles over Voronoi method. After the neighbourhood identification, we collect the statistics of non-affine deformation of the neighbourhood of individual particles in all simulated configurations.

\section{Results}
\subsection{Statistics of local stress: curious {\em pure} shear fluctuations across phase transition}
\begin{figure*}[h!]
    \includegraphics[width=0.99\textwidth]{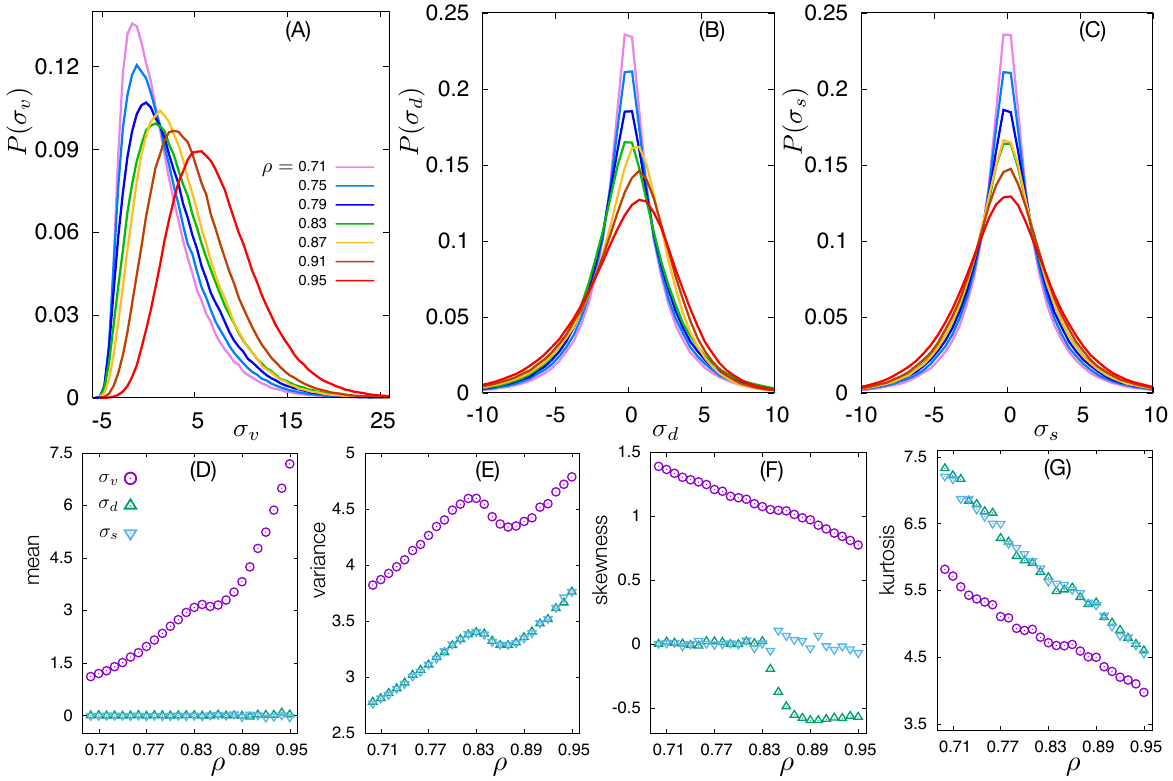}
\caption{\label{fig:strs}
{\bf Statistics of local stress:} The distributions of (A) volumetric stress ($\sigma_{v}$), (B) deviatoric stress ($\sigma_{d}$) and (C) shear stress ($\sigma_{s}$) are plotted for a set of densities. The position of $\text{max.}(\sigma_{v})$ shifts from negative to positive values across the liquid to solid transition through zero value marking the beginning of coexistence phase. For both deviatoric and shear stresses, the distributions are of Poisson type. While $P(\sigma_{s})$ is always symmetric about zero, $P(\sigma_{d})$ is only so in liquids and becomes skewed towards positive values as the system enters coexistence phase. Four primary moments of these distributions: (D) mean, (E) variance, (F) skewness and (G) kurtosis, are presented next as a function of density. Mean value of $P(\sigma_{v})$ or {\em bulk} pressure shows expected changes across the phase transition. Note that mean values of $P(\sigma_{d})$ is always zero even if it becomes asymmetric for high densities. Interestingly, the variance of all these distributions seem more sensitive to the transition than their mean values and show same qualitative behaviour across the transition. The skewness of $P(\sigma_{d})$ captures the liquid-solid coexistence most clearly showing a step-like jump to a fixed negative value in crystalline state, starting from a zero value in the liquids. The skewness of $P(\sigma_{v})$ decreases monotonically with increasing $\rho$ and approaches zero as its kurtosis also decreases monotonically to $\text{value}=3$ indicating the Gaussian distribution as expected in a perfect harmonic solid.
}
\end{figure*}

We begin by examining the distribution of local and instantaneous stresses in the the simulated configuration. The distributions of all these three components of the local stress tensors are unimodal and non-Gaussian in the liquid phase. The volumetric stress distribution, $P(\sigma_{d})$, is always skewed towards positive values. In a liquid of density as low as $\rho=0.71$, the distribution peaks at negative values of $\sigma_{v}$,  but the long tail spread over positive values keeps the mean (bulk pressure of the system) at a non-zero positive value. This maximum value gradually lowers and its position shifts to positive values of $\sigma_{v}$ as the density increases. Note that, $P(\sigma_{v})$ peaks at $\sigma_{v}=0$ for $\rho=0.83$, which marks the beginning of the coexistence phase showing an inflection in the pressure-density relation. The other two stresses, $\sigma_{d}$ \& $\sigma_{s}$, are not supposed to contribute to the bulk stress response of any equilibrium system. As expected, the distribution of deviatoric stress or {\em pure} shear, $P(\sigma_{d})$, peaks at the zero value of the corresponding stress and falls of symmetrically in a Poissonian fashion. Curiously, $P(\sigma_{d})$ becomes skewed as the system enters the coexistence phase ($\rho>0.83$) in such a way that its mean value remains zero for all densities. The distribution of shear stress, $P(\sigma_{s}$), is also of Poisson type but always symmetric about $\sigma_{s}=0$ unlike the pure shear case.

Studying the higher moments of the stress distributions reveal a few more important features about the local stress response of the system across the phase transition. For example, the second moment or {\em variance} appears to be more sensitive to the transition than the first moment or {\em mean} values. In fact, the variances of $\sigma_{d}$ and $\sigma_{s}$, whose mean values are insensitive to the transition, show two prominent inflection points marking the coexistence phase in a way, qualitatively similar as the mean and variance of $\sigma_{v}$. Most interestingly, the third moment or {\em skewness} of pure shear appears to change as a {\em perfect} order parameter which is zero valued in the liquid state and has a finite non-zero constant value in the crystalline phase. The fourth moment or {\em kurtosis} of the stress distribution are positive valued referring to their non-Gaussian nature as mentioned earlier. Note that, a Gaussian distribution will have a $\text{kurtosis}=3.0$ which $\sigma_{v}$ tends to approach monotonically as the density increases. While this is expected, this can be used to identify the {\em harmonic limit} at a given temperature.

\subsection{Statistics of local non-affine deformation: {\em Three} structural population
}
\begin{figure*}[h!]
    \includegraphics[width=0.99\textwidth]{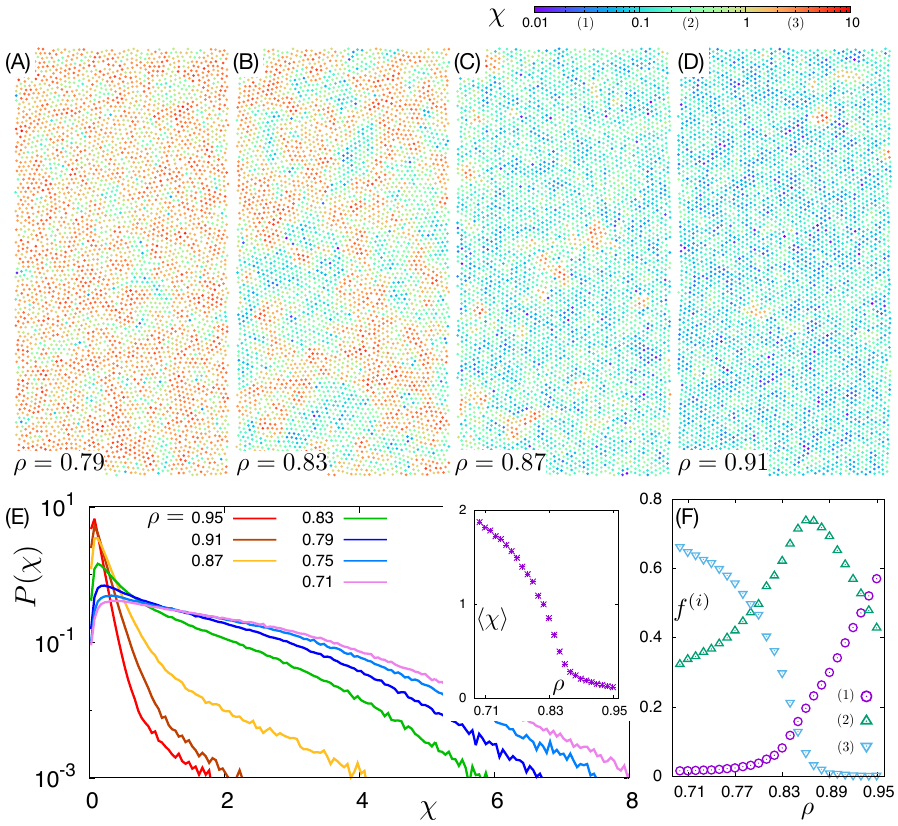}
\caption{\label{fig:nap}
{\bf Spatial map of non-affine parameter ($\chi$) and its statistics.} Particles within a small part of the simulation box are colored according to their $\chi$ values for different densities, (A) $\rho=0.79$, (B) $\rho=0.83$, (C) $\rho=0.87$ and (D) $\rho=0.91$. In the liquid state (A), most particles have {\em large} $\chi$ which gradually decreases through the phase coexistence region, (B) \& (C). Large $\chi$ regions appear only as {\em small isolated islands} in the crystalline phase (D) where the rest of the particles have low values of $\chi$. (E) The non-affine fluctuations ($0.0<\chi<1.0$) are distributed in an exponential way within a solid ($\rho=0.95$) where even larger non-affine values ($1.0<\chi<2.0$) seem to appear with yet another exponential form. This second exponential tail becomes longer as the system enters coexistence phase ($\rho=0.87$) and gets stretched to even larger values as it leaves the liquid-solid coexistence ($\rho=0.83$). In the liquid phase, the distribution of non-affine fluctuations can be best described as a stretched exponential form. ({\em Inset}) shows the ensemble average of this fluctuation, $\langle\chi\rangle$ as a function of $\rho$. $\langle\chi\rangle$ decreases monotonically in liquid phase with increasing $\rho$, drops off sharply across the phase coexistence and asymptotically approaches to zero in the solid phase as the solid approaches its harmonic limit with increasing density. {\em Three} structural populations are identified based on the $\chi$ values: (i) $\mathcal{P}_1:\chi<0.1$, (ii) $\mathcal{P}_2: 0.1\le\chi\le1.0$ and (iii) $\mathcal{P}_3: \chi>1.0$. Number fraction $f^{(i)}$ of particles belonging to each population as a function of $\rho$ is plotted in (F).
}
\end{figure*}

Next, we move to study the local deformations occurring in the system across phase transition in detail. We find that, in both liquids and solids, non-affine fluctuations are spatially heterogeneous and appear in a broad range of values: $0.01\le\chi\le10.0$. While {\em large} fluctuations ($\chi\ge1.0$), dominate in the liquid phase, it decreases gradually with increasing density and most of the particles in the solid phase show very little non-affine fluctuations ($\chi\le0.1$). In a typical crystalline system ($\rho=0.95$), the probability $P(\chi)$ shows a linear dependence with $\chi$ when plotted in a log-linear scale indicating an exponential distribution of non-affine fluctuations for $\chi\le1.0$. Even larger values of $\chi (>1.0)$ are rare in solids but do appear in small patches following another exponential form. The mean non-affine fluctuation, $\langle\chi\rangle$, appears to reach zero value as the density is increased further. Note that even in an ideal harmonic net, pure thermal agitation may result into non-affine fluctuations. As the density is decreased $\langle\chi\rangle$ increases slowly as the exponential cut-off of the distribution shifts to larger values of $\chi$. An abrupt increase in $\langle\chi\rangle$ occurs as the system enters coexistence phase ($\rho=0.87$) where the exponential tail of $P(\chi)$ becomes broader and the large non-affine fluctuations start to become significant. Upon further decrease in density, the exponential becomes stretched and large patches of high-$\chi$ regions appear in the system. In the liquid phase, such patches cover nearly the entire system and low-$\chi$ patches become rare. Motivated by such collective appearance of non-affine fluctuations, we define {\em three} structural populations: (i) $\mathcal{P}_1:\chi<0.1$, (ii) $\mathcal{P}_2: 0.1\le\chi\le1.0$ and (iii) $\mathcal{P}_3: \chi>1.0$, to capture the non-affine fluctuation within the coexistence phase where the presence of all these three populations are significant. Populations $\mathcal{P}_1$ and $\mathcal{P}_3$ are dominant in crystalline and liquid phases, respectively. $\mathcal{P}_3$ is present at all densities and becomes dominant within the coexistence phase. In the next section, we investigate the interrelation between each of these structural populations and the local stress.


\subsection{{\em Local} mutual information between stress and non-affine deformation
}
\begin{figure*}[h!]
    \includegraphics[width=0.99\textwidth]{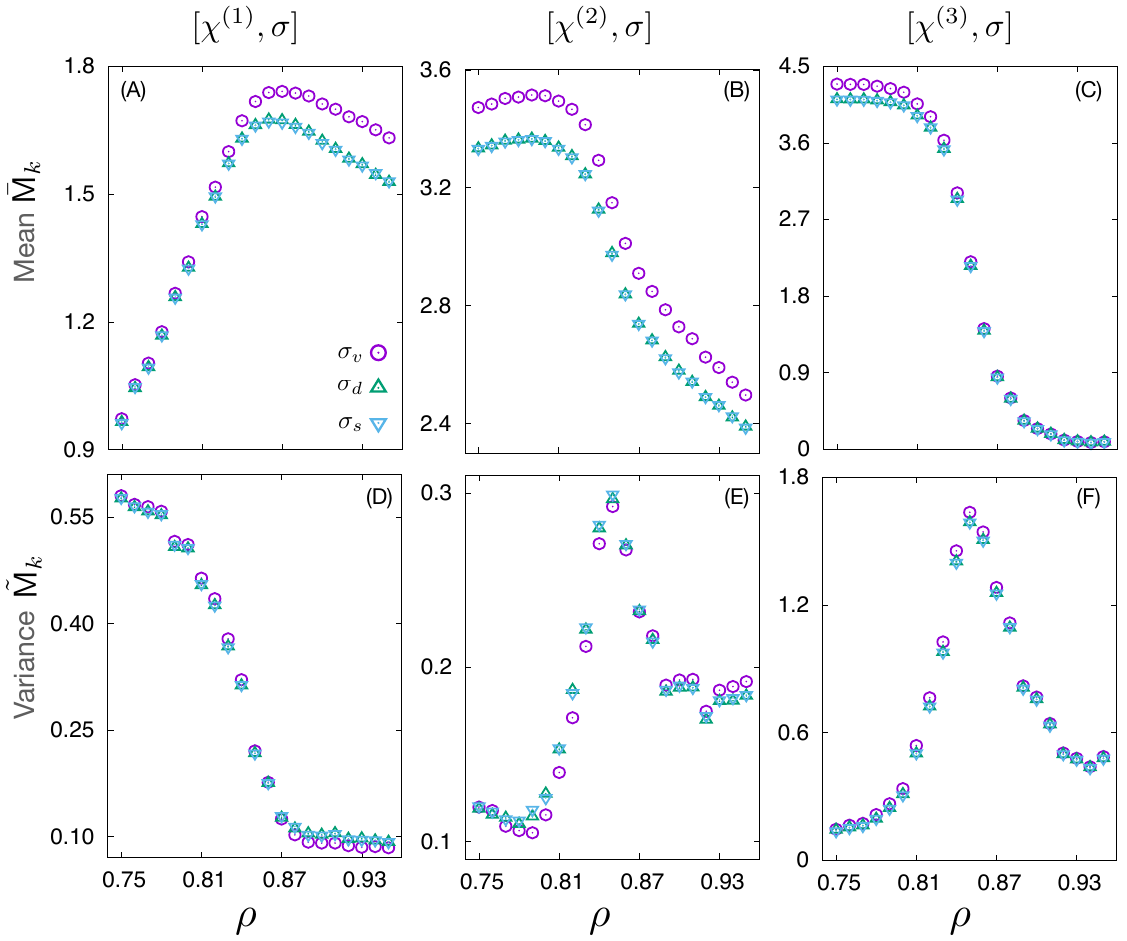}
\caption{\label{fig:localInfo}
{\bf Basic statistical properties of local stress-deformation dependence:} Local mutual information (MI) is used to quantify the dependence between virial stress invariants and non-affine deformation. ({\em see text for the definition}) Mean values of this MI, computed separately for three structural populations are plotted in (A)-(C). Note that, as a function of density, the variations for two shear components are exactly the same for each population. The pressure component shows either identical or qualitatively same variation. Respective variances for each population are plotted in (D)-(E). 
}
\end{figure*}
While a strong interdependence between local stress and non-affine deformation is intuitively very much expected, there is no formal measure to quantify the same. Here, we quantify the degree of this interdependence by the {\em local} mutual information (MI) between these two field variables. Specifically, we first divide the whole system into square sub-blocks of size $1.5\sigma$ (roughly the size of the first nearest neighbour shell)and compute MI for the $k$-th block using the standard form: $\mathsf{M}_k(\mathsf{X},\mathsf{Y}) = \mathsf{H}_k(\mathsf{X},\mathsf{Y}) - \mathsf{H}_k(\mathsf{X})-H_k(\mathsf{Y})$, where $\mathsf{X}$ and $\mathsf{Y}$ are two random variables (RVs), local stress and local non-affine deformation, in this case. $\mathsf{H}_k$'s are the block Shannon entropies: $\mathsf{H}_k=-\sum_{i\in k}P_i\ln P_i$ with $P_i$ being the probability of appropriate RV(s). With this definition, physical interpretation of MI is straightforward. A zero value of MI would ensure independence of two RVs while any non-zero value is indicative of their mutual dependence. Stronger will be the dependence as the MI values appear larger.

First, we compute the MI between each of three components of stress tensor and each of the three non-affine populations for all simulated systems. In Fig., we present their basic statistical properties, namely, their mean values $\bar{\mathsf{M}}_k=(1/n_k)\sum_k \mathsf{M}_k$ and their variances $\tilde{\mathsf{M}}_k=(1/n_k)\sum_k(\mathsf{M}_k-\bar{\mathsf{M}}_k)^2$, where $n_k$ are the total number of blocks. We note that, for each of the structural population, these properties of MI between the deformation and different stress invariants show similar qualitative behaviour when they are not identical. Recall that both of shear stresses do not contribute to any bulk response, yet there fluctuations are found to be intriguingly related to the local deformation field. 

For the structural population $\mathcal{P}_1$ ($\chi\le0.1$), $\bar{\mathsf{M}}_k$ varies within a small range of non-zero values. For all three stress invariants, these are identical in the liquid phase and grows as a function of increasing density. MI for local pressure starts to differ from the two shear stresses, for which $\bar{\mathsf{M}}_k$s are identical, as the coexistence regime starts at $\rho=0.83$. All three of them reaches their maximum at $\rho=0.87$ and start to decrease as the system becomes crystalline with further increase in density. However, for all densities, $\tilde{\mathsf{M}}^k$s show identical sigmoid-like behaviour for all three stresses: starting from a high value, they drop off rapidly across the coexistence regime and settle down to a small value in the solid phase. From this observations, we infer that stress-deformation dependence for small non-affine fluctuations is finite but overall very small.

The stress-deformation dependence is naturally more evident for populations $\mathcal{P}_2$ and $\mathcal{P}_3$ which consists of intermediate and large non-affine deformations, respectively. For both of these populations, $\bar{\mathsf{M}}_k$s are much larger in liquid phase than in the solids, which is the opposite scenario of population $\mathcal{P}_1$. This large mean values of MI start to decrease rapidly as the system approaches the coexistence phase. This drop in mutual dependence is sharp for population $\mathcal{P}_3$ as it reaches zero value in the solid phase while in population $\mathcal{P}_2$, it is a finite positive quantity and continues to decrease with increasing system density. Most interestingly, $\tilde{\mathsf{M}}_k$s show a sharp cusp-like feature as a function of density, peaking at $\rho=0.85$ for both of the structural populations. 

\begin{figure*}[h!]
    \includegraphics[width=0.99\textwidth]{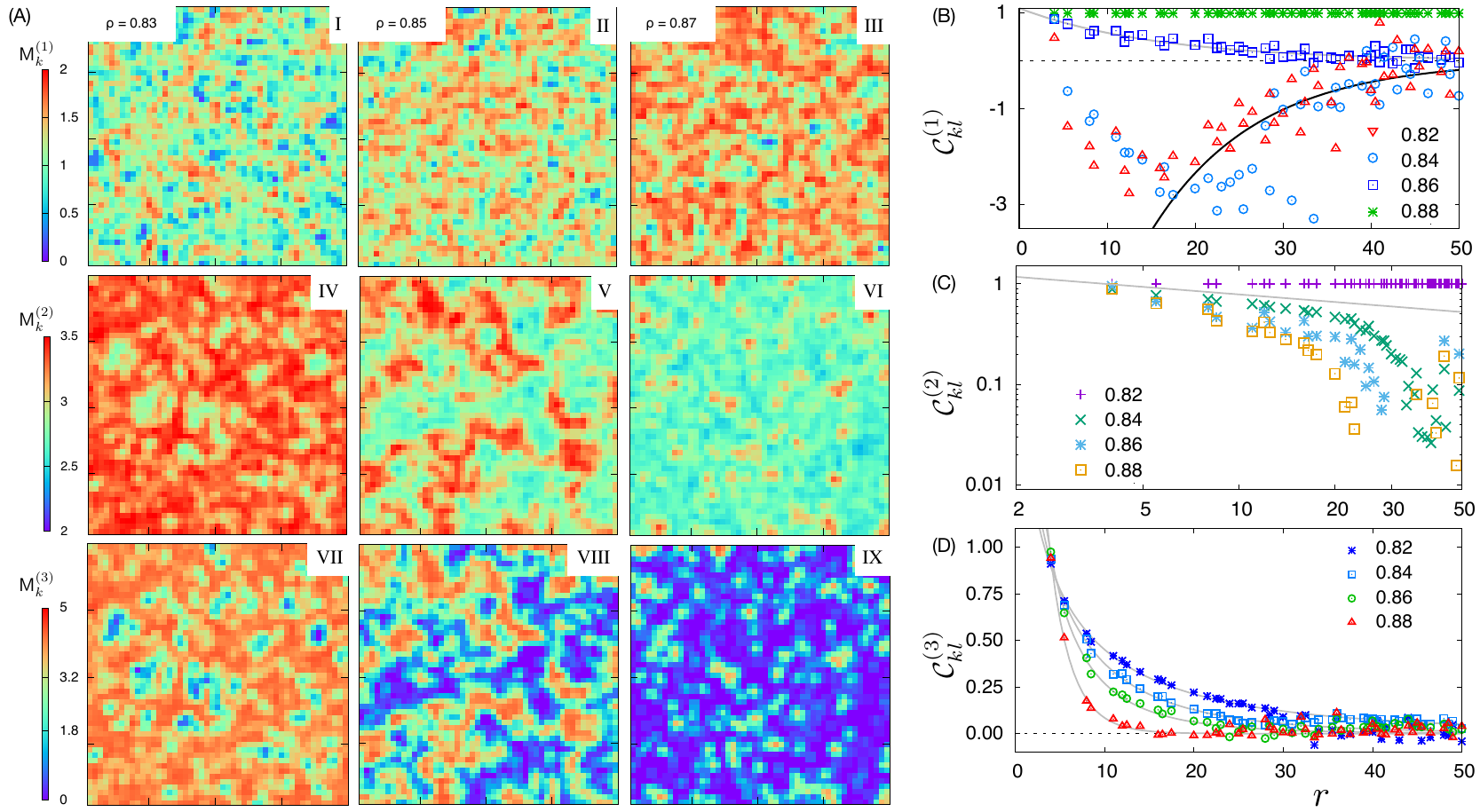}
\caption{\label{fig:splInfo}
{\bf Spatial correlation of local stress-deformation fluctuation:} (A) Spatial map of local stress-deformation MI for structural populations $\mathcal{P}_1$ (I-III), $\mathcal{P}_2$ (IV-VI) and $\mathcal{P}_3$ (VII-IX), showing heterogeneous droplet feature. Spatial correlation of this dependence is plotted separately for the three populations in (B), (C) and (D). For population $\mathcal{P}_1$, the correlation goes from negative to positive for liquid to solid transition as denoted by two solid lines of opposite exponential fitting. An algebraic decay of correlations is observed for population $\mathcal{P}_2$ across solid-liquid transition. The solid line marks a decay of $r^{-1/4}$. Spatial correlations in population $\mathcal{P}_3$ is always exponential with the cutoff distance increasing monotonically as a function of decreasing density.  
}
\end{figure*}
Second, we investigate the spatial extent of the stress-deformation dependence by drawing a visual map of the same. As we have already noted that the MI between non-affine deformation and all three stress invariants are practically the same, we can safely choose any one of them for our investigation. In Fig.\ref{fig:splInfo}, we show the spatial map of MI between $\sigma_{d}$ and $\chi$ for different structural populations across the phase coexistence region. We find that the high-MI valued blocks always appear as spatially localized {\em droplets} in the system. Within the coexistence phase ($\rho=0.85$), where all three structural populations are present sufficiently, the droplets showing {\em high} dependency between stress and deformation seem to form percolating network, individually, in all of them. For $\rho=0.83$ and $\rho=0.87$ where the presence of population $\mathcal{P}_1$ and $\mathcal{P}_3$ are minimal respectively, such high-MI droplets naturally appear in isolation, yet spatially ramified way. The situation reverses as these populations become significant as a function of density and high-MI droplets span the whole system. 

To quantify further, we compute the two-point correlation as follows:
\begin{equation}
\mathcal{C}_{kl}=\frac{\langle (\mathsf{M}_k-\bar{\mathsf{M}}_k)(I^l-\bar{\mathsf{M}}_l) \rangle}{\bar{\mathsf{M}}_k\bar{\mathsf{M}}_l}
\end{equation}
where $(k,l)$ are the indices of an arbitrary pair of blocks. $\mathcal{C}^{(i)}_{kl}$ would denote the spatial correlation for the $i$-th structural population. Curiously, $\mathcal{C}^{(1)}_{kl}$ reveals a negative spatial correlation for $\rho=0.83$ among the stress-deformation droplets within population $\mathcal{P}_1$ which appear with a very low number fraction in the system. At this density, droplet correlation is maximum and constant ($\mathcal{C}^{(2)}_{kl}=1.0$) across the system for population $\mathcal{P}_2$ which consists of most the particles in the system. In population $\mathcal{P}_3$, the second dominant population, the correlation $\mathcal{C}^{(3)}_{kl}$ is positive but dies out exponentially as a function of distance. For a slight increase in density, at $\rho=0.84$, $\mathcal{C}^{(1)}_{kl}$ remains same, $\mathcal{C}^{(3)}_{kl}$ is qualitatively same but decays faster and $\mathcal{C}^{(2)}_{kl}$ shows an algebraic decay. As the density is increased further, at $\rho=0.86$, $\mathcal{C}^{(2)}_{kl}$ and $\mathcal{C}^{(3)}_{kl}$ are qualitatively same, decaying now over even shorter length scale and $\mathcal{C}^{(1)}_{kl}$ becomes positive and exponentially decays to zero at large distance. When the system exits phase coexistence, population $\mathcal{P}_1$ spans the system and so does $\mathcal{C}^{(1)}_{kl}$, at $\rho=0.88$. $\mathcal{C}^{(3)}_{kl}$ and $\mathcal{C}^{(2)}_{kl}$ decays quickly over a distance around $10\sigma$ and $20\sigma$, respectively, at this density.

\section{Discussion}
In summary, we have compiled a spatial map of heterogeneous local stress-deformation coupling in an equilibrium two-dimensional system across the liquid-solid transition, one well-known example of global change in mechanical behaviour of materials. Statistics of local stress invariants and non-affine deformation in the simulated configurations are first studied independently. For the local stress, it is observed that the distributions of local stress components are particularly sensitive to the phase transition. Notably, the perfectly symmetric distribution of pure shear in the liquid phase is found to become skewed at the onset of transition to reach a constant skewness in the solid phase while maintaining a zero mean value throughout. For the local deformation, three structural populations are identified based on the values of non-affine fluctuation of the immediate neighbourhood of the particles. While very large and very small deformations are found to be rare in the solid and liquid phases, respectively, intermediate deformations are present in both phases and are most dominant in the coexistence phase. 

The degree of dependency between stress and non-affine deformation is then defined in terms of the mutual information of these two field variables. Further study reveals that such dependency always occurs in the system in a localised and spatially correlated way. While the length scales of these correlations are dependent on the abundance of respective populations, their function behaviours are different for different populations. Interestingly, near phase transition, the stress and intermediate non-affine deformation dependency shows an algebraic decay in spatial correlation, one characteristic feature of two dimensions. For other rare deformations, their dependency with stress decays exponentially in space and the dominant deformations are always strongly dependent on local stress with their correlation spanning across the whole system.

Motivated by these findings, we propose the stress-deformation dependency as a new collective variable to understand the microscopic mechanism of mechanical response of materials. At its current state, this description, in spirit, complements the ongoing machine learning efforts~\cite{Liu:2015, Liu:2016, Liu:2017, Manning:2020} to quantify the {\em softness} in glassy materials. In fact, the dependency matrices along with their well behaved correlation functions might be useful to train a machine algorithm faster for an investigation of nucleation or glass transition mechanism. As we have noticed that the mutual information is insensitive to the different stress invariants, this collective variable might be useful to compare the mechanical response of resulting from different protocols and understand the underlying mechanism. 

These equilibrium results has yet another fundamental aspect. Since the equilibrium system is under no external perturbation, the local stress and associated deformations are the natural building blocks for any mechanical response theory to build upon. The coupling between them quantified by this new collective variable is then an efficient identifier of an elementary localized excitation which reliably captures the global mechanical change in the system. At present, the computed statistics is sufficient for a mean field description and further detailed analysis needs to be be performed for writing down a formal theoretical framework. However, this descriptor, being rooted in the information theory, can be applied to any material irrespective of their chemical composition and/or conformational details, as long as the distributions of the two ingredient variables are available. Further, this new description also opens up the possibility to understand the response of materials as an outcome of dynamic information exchange which is the case in many biological systems such as developing embryos, matured tissues and neuronal networks.

\bibliography{main}

\end{document}